\documentclass[fleqn, usenatbib, useAMS]{mnras}

\usepackage{graphicx}    
\usepackage{amsmath}    
\usepackage{amssymb}    
\usepackage{multicol}   
\usepackage{bm}            
\usepackage{pdflscape}    

\usepackage[T1]{fontenc}
\usepackage{ae, aecompl}
\usepackage{newtxtext, newtxmath}

\title[Cooling of adiabatic discs in merging BBH]{The effect of cooling on the accretion of circumprimary discs in merging supermassive black hole binaries}

\author[C. Fontecilla, G. Lodato \& J. Cuadra]{
{Camilo Fontecilla}$^{1}$\thanks{E-mail: \href{mailto:cfonteci@astro.puc.cl}{cfonteci@astro.puc.cl}}, {Giuseppe Lodato}$^{2}$ and {Jorge Cuadra}$^{1,3}$
\\
$^{1}$ Instituto de Astrof\'isica, Pontificia Universidad Cat\'olica de Chile, Av. Vicu\~na Mackenna 4860, Santiago, Chile\\
$^{2}$ Dipartimento di Fisica dell' Universit\`a degli Studi di Milano, Via Celoria 16, Milano, Italy\\
$^{3}$Departamento de Ciencias, Facultad de Artes Liberales, Universidad Adolfo Ib\'a\~nez, Av.\ Padre Hurtado 750, Vi\~na del Mar, Chile
}

\date{Accepted 20XX . Received 20XX}
\pubyear{2020}

\begin{document}
\label{firstpage}
\pagerange{\pageref{firstpage}--\pageref{lastpage}}
\maketitle

\begin{abstract}
At the final stages of a supermassive black hole coalescence, the emission of gravitational waves will efficiently remove energy and angular momentum from the binary orbit, allowing the separation between the compact objects to shrink. In the scenario where a circumprimary disc is present, a \textit{squeezing} phase will develop, in which the tidal interaction between the disc and the secondary black hole could push the gas inwards, enhancing the accretion rate on to the primary and producing what is known as an electromagnetic precursor.
In this context, using 3D hydrodynamic simulations, we study how an adiabatic circumprimary accretion disc responds to the varying gravitational potential as the secondary falls onto the more massive object. We included a cooling prescription controlled by the parameter $\beta=\Omega t_{\rm cool}$, which will determine how strong the final accretion rate is: a hotter disc is thicker, and the tidal interaction is suppressed for the gas outside the binary plane. Our main results are that for scenarios where the gas cannot cool fast enough ($\beta \geq 30$) the disc becomes thick and renders the system invisible, while for $\beta \leq 10$ the strong cooling blocks any leakage on to the secondary's orbit, allowing an enhancement in the accretion rate two orders of magnitude stronger than the average through the rest of the simulation.
\end{abstract}

\begin{keywords} 
accretion, accretion discs -- methods: numerical -- black hole physics -- hydrodynamics -- gravitational waves 
\end{keywords}



\section{Introduction}

In the center of our galaxy, Sgr A*, a Super Massive Black Hole (SMBH) with a mass of four million suns lays almost dormant. From the theoretical point of view and by indirect evidence \citep{Ferrarese2000, Kormendy2013, Reines2014}, we know that this kind of black holes, with masses from $10^5$ to $10^9$ solar masses, inhabit the centre of most, if not all, the more massive galaxies in our universe.

While a black hole itself does not emit light, the gravitational interaction with its close vicinity can make it luminous. The current accretion theory explains the energy observed and the shape of the spectra in the centres of active galaxies as the infall of material from an accretion disc into a point like mass \citep{Frank2002}. At higher redshift, the amount of gas in the galaxies alongside with their interactions and merger increases. There is compelling evidence \citep{Begelman1980, Volonteri2003} that a binary system of SMBHs can form when their host galaxies merge, while simulations show that the binary separation should decrease due to different processes depending on the separation and the environmental conditions \citep{Merritt2005, Colpi2011, Mayer2019}. In the observations, systems at different stages of evolution have been identified \citep{Colpi2011} with a clear bias toward larger separations. The absence of systems at distances smaller than $\mathcal{O}(1)$ parsec represents a challenge to both observational and theoretical astrophysics. Different explanations have been proposed \citep{Shannon2015} to solve this problem: on the one side; it could be that the SMBHs separation stalls at some point in the galaxy merger and without a continuous inflow of material feeding the black holes they eventually become undetectable. The opposite scenario considers that the black holes continue shrinking. However, the short timescales of the processes at small distances make the detection of a system in this stage extremely unlikely \citep{Begelman1980}.

Extensive theoretical work has been already done studying the SMBH migration at different scales due to different processes: in order for the binary to merge in less than a Hubble time, we need a combination of mechanisms that extract energy and angular momentum from the system. 
At large scales, dynamical friction dominates \citep{Governato1994, Milosavljevic2001, Escala2005}, while at parsec scales gas clouds \citep{Goicovic2017}, the presence of a circumbinary accretion disc \citep{Milosavljevic2003, Cuadra2009, Roedig2011} or the scattering of stars \citep{Yu2002} can efficiently reduce the black hole distance. At even smaller distances, a combination of a disc and the gravitational wave (GW) emission (which becomes dominant at separations of $\mathcal{O}(100)$ Schwarzchild radii ($\mathrm{R_S}$) \citep{Lodato2009, Kocsis2012, Armitage2002, Fontecilla2019} will produce the final merger \citep{Peters1964}.

GWs have been already observed in the stellar-mass regime by LIGO \citep{Abbott2016}. However, it is not until LISA becomes operational that we will be able to observe the emission from a SMBH binary system merging \citep{eLISA2013}. 
The technique used to detect GWs is not enough by itself to determine precisely the location of the merger. For this reason, an electromagnetic (EM) counterpart is crucial to know the position of the event \citep{Centrella2010}. In the case of a BH-BH system, this counterpart comes from the material in the vicinity minutes prior the final merger \citep{Armitage2002, Chang2010, Tazzari2015, Fontecilla2017}.
As such, this counterpart works as an alert, or precursor, for an imminent burst of GWs.
For this enhancement to occur, the environment needs to be such that material can be funnelled on to the central region and form, due to the varying gravitational potential of the binary, a truncated circumbinary disc. Then, gas should migrate due to viscous process and cross the hollow cavity surrounding the SMBHs, to finally become part of the circumprimary disc.

Even if material manages to survive the whole binary evolution, for a merger to be detectable, the enhancement in the luminosity of the binary system needs to surpass the emission of the host galaxy (or the quiescent accretion disc emission).
The tidal interaction between the gas in the inner disc and the shrinking binary could produce this enhancement. Due to the GW emission, the migration timescale of the black holes eventually becomes shorter than the viscous timescale of each accretion disc in the system: around $a \sim 10^2 \mathrm{R_S}$ a \textit{first decoupling} occurs: the circumbinary disc is left behind, and the migration becomes entirely dominated by the GW emission \citep{Armitage2002, Lodato2009}. At even smaller distances ($a \sim 40 \mathrm{R_S}$) the inner disc strongly couples with the gravitational potential of the binary and is pushed inwards in the \textit{squeezing} phase, potentially enhancing the accretion rate on to the primary \citep{Chang2010, Tazzari2015, Fontecilla2017}.
For this process to be efficient and produce the precursor, the disc-binary coupling needs to be maintained: the closer the gas is to the midplane, the stronger is the gravitational interaction. For this reason, the thickness of the disc plays a significant role in the enhancement of the accretion rate \citep{Fontecilla2017}. Depending on the scale height, gas can cross the secondary's orbit outward following horseshoes trajectories \citep{Baruteau2012}, or accreted on to the primary \citep{Cerioli2016}.
In this context, \citet{Cerioli2016} made 3D hydrodynamical simulations of non-equal mass SMBHB with isothermal discs in the squeezing phase. 
Later on,  \citet{Pereira2019} demonstrated that this luminosity enhancement is robust with respect to the possible misalignment between the disc and the binary orbit, so that even discs that are misaligned by several degrees still show a non-negligible enhancement in the accretion rate, above the Eddington limit. Here, we want to test the robustness of this scenario with respect to changes in the gas thermodynamics. Indeed, instead of using an isothermal equation of state (EoS) as in previous works, which implies that the disc can cool down efficiently in order to maintain a fixed scale height $H / R$, here we use an adiabatic equation of state with an added model for the cooling. The main objective is to theoretically study how this kind of accretion discs behave in a tidally dominated environment. 
 
Our result in \citet{Fontecilla2017} motivates this change in the EoS. Using a 1D code,  we found a \textit{second decoupling} between the inner disc and the secondary black hole: at around $a \sim 20 \mathrm{R_S}$, the gravitational influence of the smaller companion is suppressed by a combination of tidal heating and inefficient cooling through the disc. As the temperature increases, the disc becomes thick, which could allow it to survive the SMBH merger instead of being fully accreted by the primary. 
To a first approximation, we can use \citet{Pereira2019}'s scenario and re-interpret it as one with an aligned disc of thickness $H \sim \arctan(i) \times R$, from their results, an expected scale height of $H / R \sim 0.3$ will be enough to suppress the squeezing effect and not produce a precursor.
 
In the following sections, we discuss our implementation, assumptions and how our work fit in the already existing literature, what are the improvement that can be done to model this kind of system accurately, and our main conclusions in this scenario.

\section{System properties}
\subsection{Late stage Merging SMBHBs}

Independent of the nature of a binary system, the presence of a secondary object with mass $M_\mathrm{s}$ produces a gravitational potential that varies over time \citep{Lin1979}. The overall geometry of the system and the extension of the accretion discs will be determined by the mass ratio $q = M_\mathrm{s} / M_\mathrm{p}$, where $M_\mathrm{p}$ is the primary, more massive member of the binary \citep{DOrazio2016}. The disc properties also play a role in its size and the inflow of material through the secondary's orbit \citep{Artymowicz1994, Ragusa2016}. For an aligned circumbinary disc, resonance effects will transfer angular momentum between the binary and the gas, truncating the outer disc at a distance around twice the black hole separation and allowing them to migrate \citep{Lin1986}.

GWs carry away angular momentum and energy from the system emitting them. In the case of a SMBH binary with a $10^8$ solar mass primary, if the separation between the black holes become smaller than $\sim 10 ^ {- 3}$ parsec, the GW emission will dominate the migration and force the black holes to merge in less than a Hubble time. Assuming a circular binary, we have \citep{Peters1964}:
\begin{equation}
  \frac{da}{dt} = - \frac{8 c}{5} q (1 + q) \left(\frac{R_S}{a(t)}\right) ^ 3, \label{eq:dadt}
\end{equation}
where $\mathrm{R_S} = 2 G M_\mathrm{p} / c ^ 2$, $G$ is the gravitational constant and $c$ the speed of light. 

As the timescale related to this migration $t_{\mathrm{GW}} = a / |da / dt| $ is proportional to the binary separation to the fourth power, the black holes speed up at the final stages of the merger, while the viscous timescale $t_\nu = 2 R^2 / (3 \nu)$ of the accretion discs stay mostly constant. Since we are interested in the behaviour of the circumprimary disc after the start of the squeezing phase, we can completely neglect the circumbinary disc.

\subsection{Accretion disc properties}

The material of a viscous accretion disc in a binary system will evolve by a combination of internal (viscosity) and external (tidal) processes. At a given radius, the gas will follow a slightly sub-keplerian angular velocity $\Omega \leq \Omega_k$ due to pressure corrections in the momentum equation. As this angular velocity decreases away from the origin, subsequent rings of material will feel friction, which transfers momentum outwards and dissipate energy. In the isothermal case, the energy is assumed to instantly leave the disc in order to maintain a fixed temperature profile, which implies a fixed sound speed and scale height.
On the other hand, in the adiabatic case, no internal energy is dissipated, and the temperature of the disc will rise, increasing the thickness and potentially changing the effects of the external (tidal) process in the disc. 

While a locally-isothermal EoS can be a good approximation for an accretion disc dominated by viscous heating, in the squeezing phase where the tidal heating becomes relevant \citep[][]{Fontecilla2017}, we need to take into account the time needed for the disc to cool down. A more correct EoS will be something in between the two extremes, for this reason, we used an adiabatic EoS and included a simplified model for the cooling \citep{Gammie2001} to explore different scenarios where the disc is more or less efficient dissipating its internal energy.

\subsection{Numerical model}

In order to model a circumprimary accretion disc in a binary system of SMBHs merging due to GW emission, we use the Smoothed Particles Hydrodynamics (SPH) code \texttt{PHANTOM} \citep{Price2018}. To be able to make a direct comparison with  \citet{Cerioli2016}, 
we consider an unequal mass binary of SMBH with a mass ratio $q = 10 ^ {- 3}$ (qualitatively very  similar to the $q \approx 10 ^ {- 2}$ ratio used in earlier works \citep{Armitage2002, Baruteau2012}). In code units, $M_\mathrm{p} = 1$ and we set $ G = c = 1$ so the gravitational radius $\mathrm{R_g} = \mathrm{R_S} / 2 = 1$ and $t_0 = \mathrm{R_g}/c =1$. 
The circumprimary disc is modeled using $N = 5\times 10^5$ SPH particles, since this resolution is enough to capture the more relevant features of the disc evolution and is not computationally prohibitive. The initial mass of the disc is $M_\mathrm{d} / M_\mathrm{p} = 10 ^ {- 8}$, and it extends from $R_{\mathrm{in}} / \mathrm{R_g} = 2$ to $R_{\mathrm{out}} / \mathrm{R_g} = 4.1$.
As explained before, we use an adiabatic EoS: $P = (\gamma - 1) \rho u$, with $\gamma = 5 / 3$ the adiabatic index, $\rho$ the gas density and $u$ its internal energy. The election of this index over $\gamma = 4/3$ is discussed in Section \ref{sec:caf}.Finally, we enable the $\beta$-cooling prescription implemented in \texttt{PHANTOM}, so that $t_{\mathrm{cool,\beta}} = \beta / \Omega_k$ \citep{Gammie2001}, and ran simulations with different values between $\beta = 10$ and $\beta = 50$.

The initial binary separation is set to $a_0 / \mathrm{R_g} = 4.75$ due to numerical limitations: as the disc is accreted, the viscosity is enhanced, which in turn depletes the disc faster \citep{Cerioli2016}. Additionally, given that the merger timescale grows as $a_0^4$, starting farther away from the primary rapidly puts us in a regime where the simulation takes too much time and computational power to be feasible.

The simulation ends at $t / t_0 = 9500$, when the SMBH separation is equal to accretion radius of the primary $R_{\mathrm{acc,p}} = 2$ plus the secondary $R_{\mathrm{acc,s}} = 0.2$. Since the GW emission completely dominates over the back reaction of the disc in the binary, the SMBH evolution is prescribed according to Equation \eqref{eq:dadt} and the gravitational effect of the binary on the disc is implemented as an external potential.

The angular momentum transport in the disc is modeled using the $\alpha$-parameter turbulent viscosity formalism $\nu = \alpha_{\mathrm{ss}} c_s H$ \citep{Shakura1973}, where $H = c_s/ \Omega_k$ is the disc thickness and $c_s$ the sound speed. The SPH code uses the following equation to mimic this viscosity parametrization:
\begin{equation}
  \alpha_{\mathrm{ss}} = \frac{\alpha^{\mathrm{av}}}{10}\frac{\langle h \rangle }{H} \label{eq:alphass},
\end{equation}
where $\langle h \rangle \propto \rho^{- 1 / 3}$ is the the azimuthally averaged smoothing length of SPH particles and $\alpha^{\mathrm{av}}$ a numerical constant in order to give $\alpha_{\mathrm{ss}} = 0.01$ at the beginning of the simulation. 

While in the isothermal case Equation \eqref{eq:alphass} is proportional only to the number of particles and the radius for a given $H/R$, in the adiabatic one $H$ is also affected by the properties of the disc, making the behaviour of the viscosity vary over time. In particular, one can show that in our numerical model $\alpha_{\mathrm{ss}}\propto H^{-2/3}$ and that thus the viscosity $\nu\propto H^{4/3}$: as the disc becomes thicker, the angular momentum transport become stronger. 

The initial condition for the disc and its temperature depend as a power law on radius and are set in order for the ratio $\langle h \rangle / H$ to be constant at $t=0$ (a uniformly resolved disc). To achive this, we follow \citet{Price2018} and set $p = 1.5$ and $q = 0.75$.

Here we present eleven simulations: a control case without secondary for each EoS, an isothermal simulation with the same conditions as in \citet{Cerioli2016}, one pure adiabatic simulation without cooling, and seven adiabatic simulations with $\beta = \{10, 15, 20, 25, 30, 40, 50\}$. The differences between each simulation are summarized in Table \ref{tab:params}.  A few additional simulations that test the robustness of our results are briefly discussed in Section \ref{sec:caf}.

No relativistic effects are included in the model except for the imposed binary orbital decay.  (also discussed in Section \ref{sec:caf}.) 

While this is only an approximation of the real behaviour of the system, it is useful to make comparisons with previous work \citep{Baruteau2012, Cerioli2016, Pereira2019}, and to show how an adiabatic disc evolves in a tidally dominated environment and what are the consequences of the increased thickness in the tidal coupling.

\begin{table}
  \caption{Top: Common simulations properties in code units, from left to right the columns are: number of particles used, time to merger, numerical viscosity parameter, scale height, disc mass, accretion radius of the primary (equal to the inner boundary of the disc) and outer boundary of the disc. Bottom: Properties that change between simulations. From left to right, the columns are: simulation number, mass ratio, initial separation, accretion radius of the secondary, Equation of State, adiabatic index and cooling parameter. Simulation $4$ is our fiducial case. A table element with the symbol "$\times$" means that this parameter is not in the simulation, while "$-$" means is the same as the fiducial case. See text for a more detailed explanation.}
  \label{tab:params}
  \begin{tabular*}{.95\columnwidth}{cccccccc} 
        \hline
    \\[-2ex]
        $N$ & $t$ &  $\alpha_{\mathrm{av}} $ & $H / R$ & $M_d$ & $R_{\mathrm{in}}$ & $R_{\mathrm{out}}$ \\ 
        \hline
    \\[-2ex]
    $5\times10^5$ & $9500$ & $0.1117$ & $0.01$ & $10^{- 8}$ & $2$ & $4.1$ \\ 
    \\[-2ex]
      \hline
      \\[1ex]
  \end{tabular*}%
\end{table}
\begin{table}
  \begin{tabular*}{.95\columnwidth}{lcccccc}     
    \hline
    \\[-2ex]
    $N^{\rm o}$ & $q$ & $a_0$ & $R_{\mathrm{acc, 2}}$ & $\mathrm{EoS}$ & $\gamma$ & $\beta$\\
    \hline
    \\[-2ex]
    1 & $\times$ & $\times$ & $\times$ & ISO & $1$ & - \\ 
    2 & $\times$ & $\times$ & $\times$ & - & - & - \\ 
    3 & - & - & - & ISO & $1$ & - \\ 4 & $10^{- 3}$ & $4.75$ & $0.2$ & AD & $5 / 3$ & $\times$\\ 
    5-11 & - & - & - & - & - & \begin{tabular}{@{}c@{}}$10,~15,~20,~25$ \\ $30,~~40,~~50$ \end{tabular} \\
  \hline   
  \end{tabular*}%
\end{table}

\subsection{Cooling prescription}

To mimic the thermodynamics of a disc which is between an isothermal and adiabatic EoS, we use the $\beta$-cooling from \citet{Gammie2001}. This prescription assumes that the cooling timescale is related to the dynamical timescale of the disc by a free parameter $\beta$. Increasing this value makes the cooling process slower.
This energy lost can be understood from the timescales involved in it. 

The cooling timescale of a disc is given by the ratio between the internal energy and the flux ($F$) through the surface $t_\mathrm{cool}=\Sigma c_s^2 / F$, where $\Sigma = \rho / 2 H$ is the disc local surface density. Hydrostatic equilibrium in the vertical direction implies that the scale height is $H = \kappa F / c \Omega^2$, with $\kappa$ the opacity. Using both expressions, the timescale in which an optically thick disc dominated by electron scattering cools down due to radiation is $t_\mathrm{cool} = \kappa \Sigma H / 2c$.
On the other hand, the material in the disc heats up by a combination of viscous dissipation ($D_\nu$) and tidal effects ($D_\Lambda$). During the squeezing phase, the tidal effect dominates over viscosity, and the gas accretion timescale is related with the binary shrinking due to gravitational waves as $|r/v_r| \simeq t_\mathrm{GW}$ \citep{Fontecilla2017}. In equilibrium, the radiation flux at the disc surface should be equal to the heating $F = D_\nu + D_\Lambda \sim D_\Lambda$. Using this considerations, we can relate the scale height of the disc $H / R$ with its ability to cool down the heating coming from tidal effects, and write:
\begin{equation}
  \frac{H}{R} = \sqrt{\frac{t_{\mathrm{cool}}}{2t_{\mathrm{GW}}}}.\label{hr}
\end{equation}

If we use the $\beta$ prescription for the disc cooling $t_\mathrm{cool} = t_\mathrm{cool,\beta} = \beta \Omega^{-1}$ in the previous equation, we obtain $H / R \sim 4 \times 10^{- 3} (R/R_{g})^{- 3 / 4}\sqrt{\beta}$ for the properties of our model. This means that, in order to get  $H / R \sim 0.01$ as we used in the isothermal case, we need a $\beta \leq 10$. In the following we will see that the proper value is closer to $\beta \sim 20$.

A different way to estimate $\beta$ comes from equating the radiative cooling timescale to the $\beta$-cooling recipe used,
\begin{equation}
  t_{\mathrm{cool}} = \kappa \Sigma \frac{H}{2c} = \frac{\beta}{\Omega} \rightarrow \beta \sim \Omega \frac{\kappa}{4 \pi R} \frac{M_\mathrm{d}}{c} \left(\frac{H}{R}\right),
\end{equation}
from which we obtain an expression for $\beta$ that depends on the physical condition of the disc. 
Using a 1 $M_\odot$ as the initial mass \citep{Tazzari2015,Fontecilla2019} and the scale height of our model ($H/R=0.01$), we can estimate which value mimics the proper cooling timescale, from which we obtain $\beta \sim 450 - 80$ at $2\mathrm{R_g} - 4 \mathrm{R_g}$.

This $\beta$ implies that, even from the beginning, the disc should be almost adiabatic. Also, it means that the thin disc used as an initial condition is not necessarily the correct approximation, since the disc should be heated as soon as the squeezing phase began.

We confirm this using Equation (\ref{hr}) with the radiative cooling timescale to calculate the expected thickness at this binary separation, obtaining $H/R \geq 1$, implying that the inner disc is already thick.
Note that, contrary to the analytical derivation in \citet{Fontecilla2017}, in our simulations we do not assume thermal equilibrium.

Nevertheless, as the disc evolves and its total mass decreases, $\beta$ should also decline, which in turn enhances the cooling and decreases the thickness of the disc. The heating from viscosity and the tidal effect turns the evolution of the cooling too complex to follow, and since the recipe used is already a simplification, we fixed $\beta$ between $10 - 50$ and tested how the simulation behaves compared with the isothermal and pure adiabatic case.

\section{Results and Analysis}

We now present the results of eleven simulations with properties detailed in Table \ref{tab:params}. We start with the cases without secondary to address the effect of the EoS (1-2), following with a comparison between the isothermal and pure adiabatic cases with a binary (3-4). Finally, we compare all cases with a secondary, to see how the cooling changes the properties of the disc.

\begin{figure}
 \includegraphics[width=.9\columnwidth]{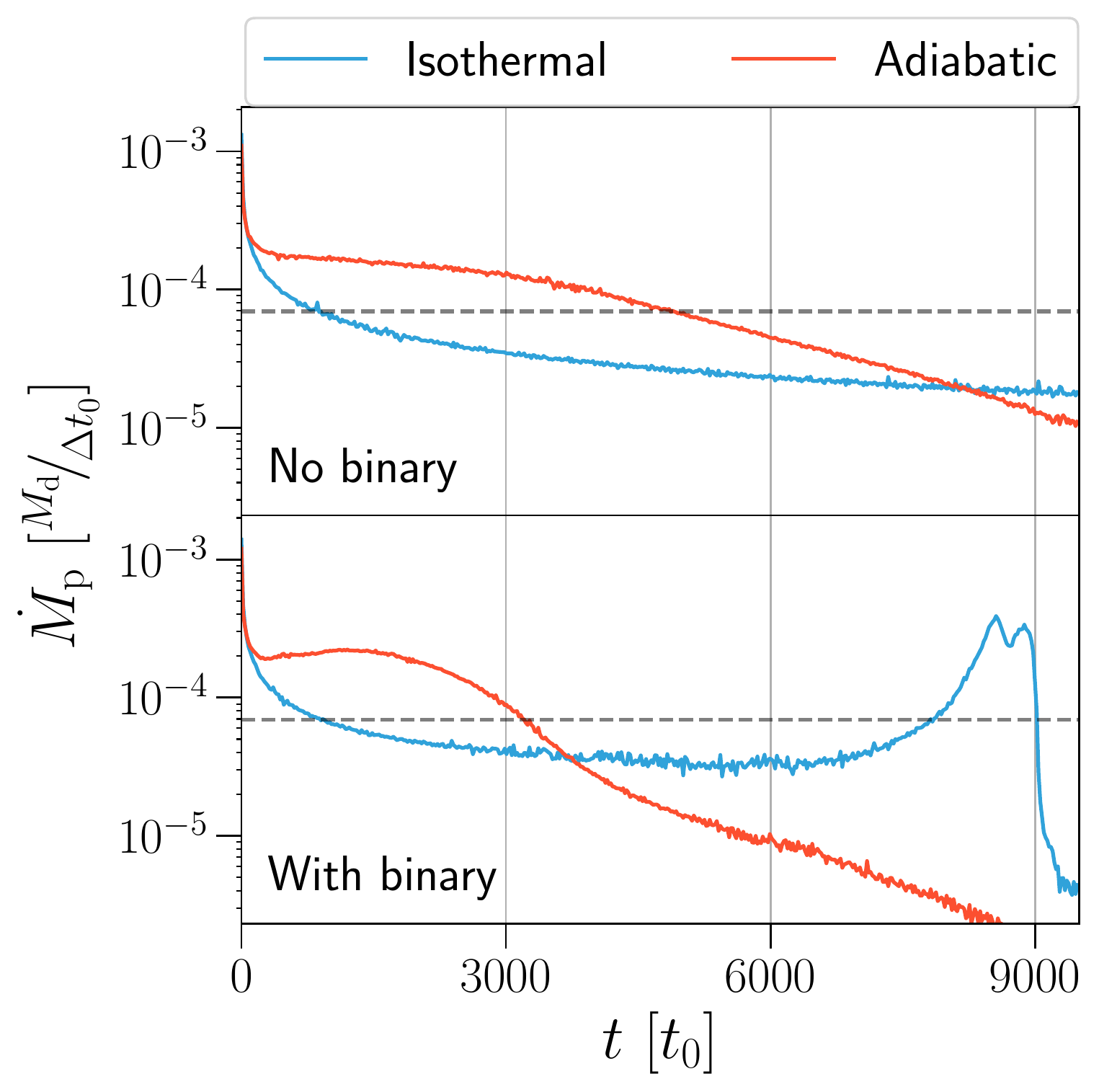}
 \caption{Top panel: accretion on to the central SMBH in the simulations without a secondary, as expected, no enhancement of the accretion occurs at late times. The blue line is the isothermal case (1) while the red is the adiabatic one (2). 
 Bottom panel: Accretion on to the primary black hole in the simulations with secondary ($3-4$), the colours follow the same trend as before. The horizontal dashed line in each plot is the Eddington accretion limit for the system. A clear difference can be seen between the two cases, with a precursor for the isothermal EoS.
 }
 \label{fig:plot1}
\end{figure}

\begin{figure*}
 \includegraphics[width=\textwidth]{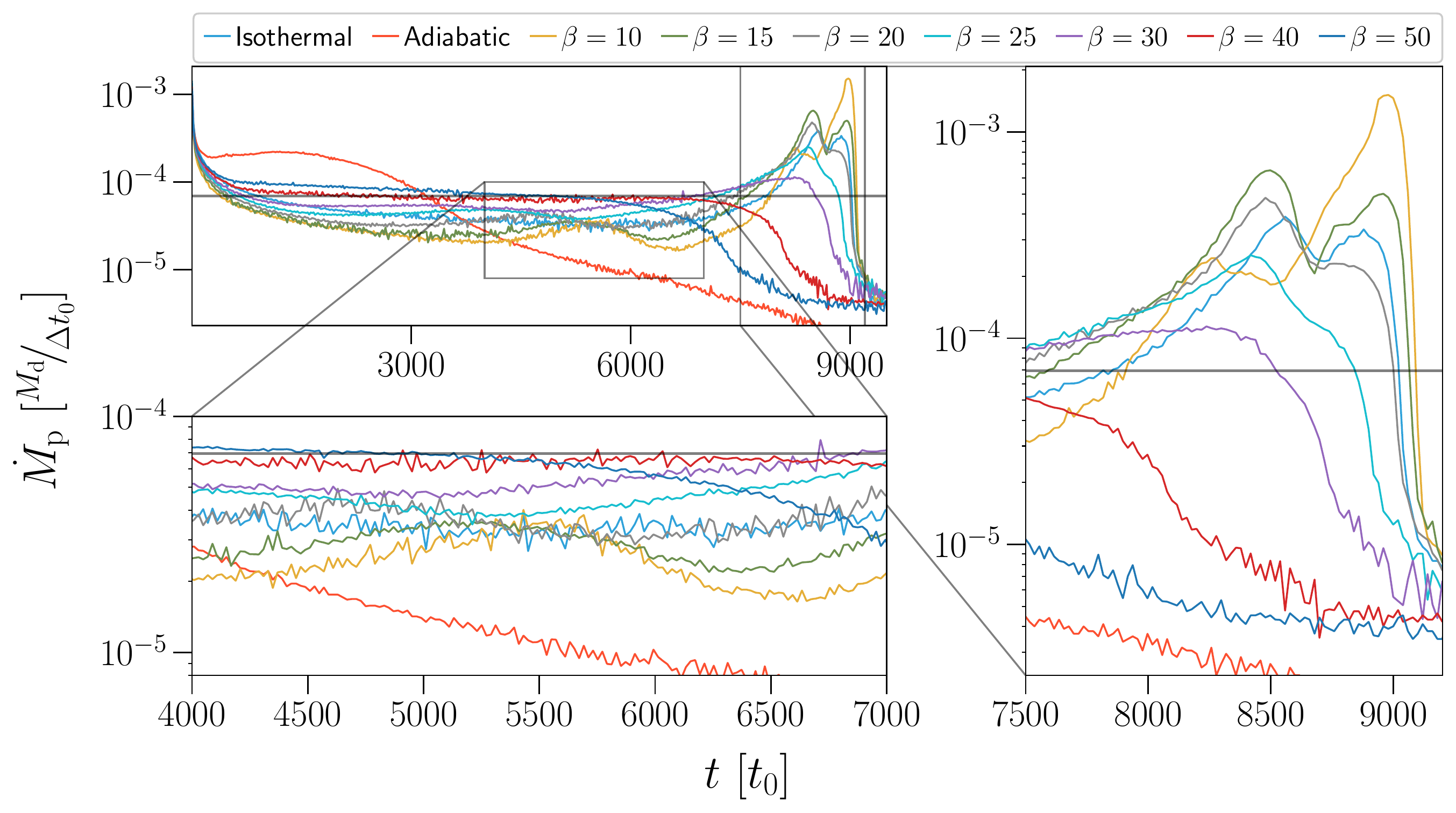}
 \caption{Top Left panel: accretion rate on to the central SMBH for all the simulations with a binary companion. Depending on how efficient the cooling is, different features arise in the simulations.
 Bottom left panel: Zoom-in in the middle of the simulations, where the  $\beta = 10$ case presents a lump which is not present for less efficient cooling.
 Right panel: Zoom-in at the end of the simulations when the precursor occurs for $\beta < 30$. Two peaks develop, and the second peak dominates for thinner discs.}
 \label{fig:plot2}
\end{figure*}

\subsection{Isothermal vs pure adiabatic}

The top panel of Figure \ref{fig:plot1} shows the accretion rate for the first (in blue, isothermal) and second (in red, adiabatic) simulations, where the disc surrounds a single black hole with mass $M_{\mathrm{BH}} = 1$. The absence of a companion manifests itself in the decaying behaviour of the accretion over time. Since the only heating factor comes from viscosity, as time goes by the disc spreads both ways and the fraction accreted by the central object decreases with the available material in the inner region.
The effect of the sharp initial condition in the surface density of the disc is visible for $t \leq 100$. After the initial transient behaviour, the adiabatic case (having a larger viscosity) presents a consistently enhanced accretion rate for most of the simulation. Only at the end, when $t \geq 8000$, the disc becomes depleted, and the inflow drops below the isothermal scenario.

On the other hand, when a secondary is present (bottom panel of the same figure), its gravitational effect will change the behaviour of the disc, either pushing the material inwards or funnelling it outside the secondary's orbit \citep{Baruteau2012}. The companion influence is evident in the adiabatic scenario, where an enhanced accretion happens in the first quarter of the simulation, and then sharply drops when most of the remnant disc spreads over a large radius where the tidal effect is no longer efficient. Instead,  in the isothermal case, the effect of the secondary is not as predominant at early times, with only a small increase in the slope of the accretion, making it somewhat constant. Only at the end, the squeezing of the inner disc allows the system to reach super Eddington accretion rates, as already discussed in \citet{Cerioli2016}.

\begin{figure*}
 \includegraphics[width=\textwidth]{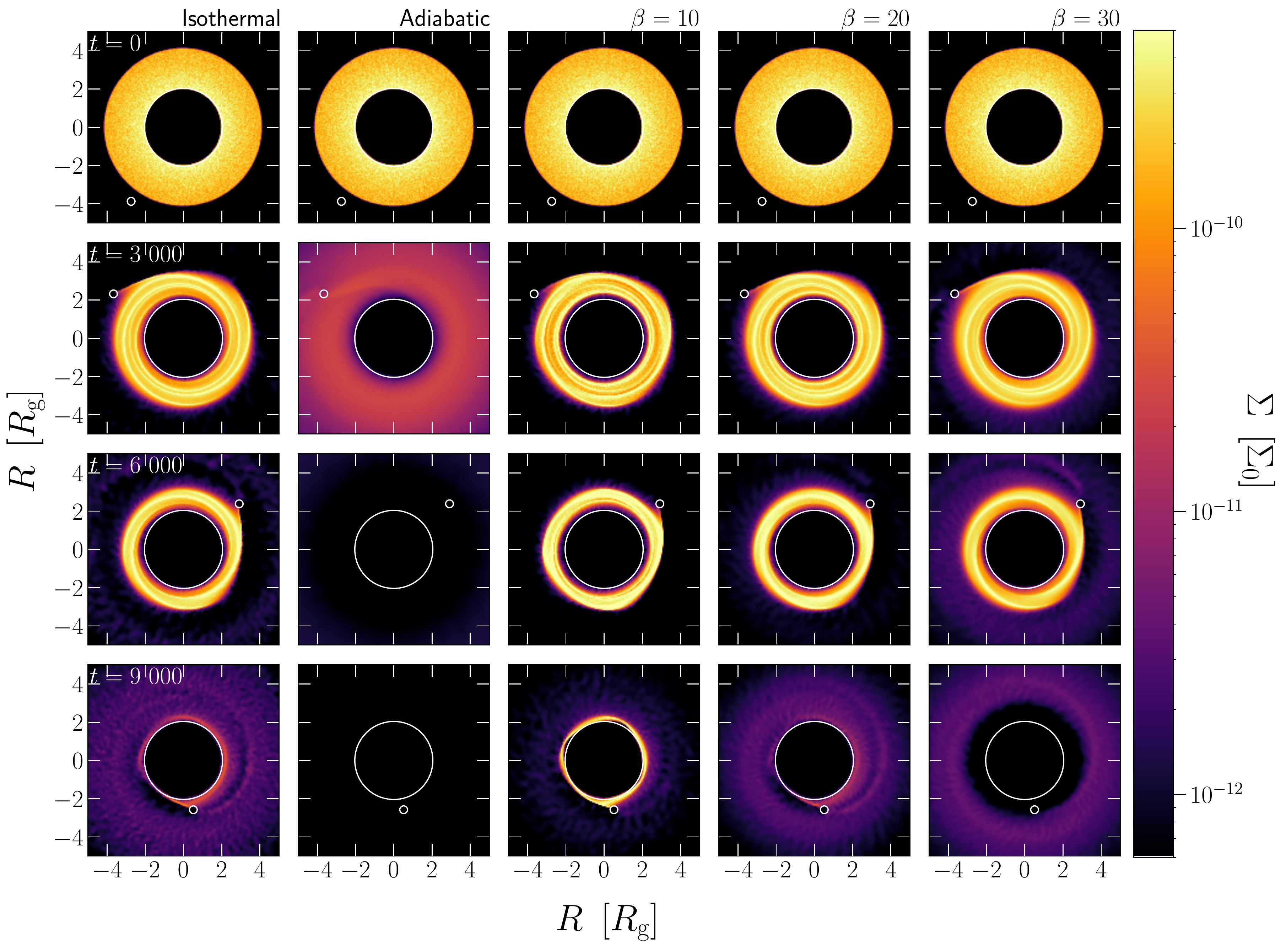}
 \caption{Surface density ($\Sigma$) in code units for five simulations, from left to right: Isothermal, pure adiabatic, adiabatic with $\beta = 10, 20, 30$. From top to bottom, initial condition ($t = 0$), after transient accretion and before lump ($t = 3000$), after lump but before final stage ($t = 6000$), final stage of the merger($t = 9000$).
 The isothermal case behaves similar to the $\beta = 20$ scenario, while a more efficient cooling allows a dense ring of material survive until the end, while a less efficient depletes the inner disc and push the material outside the secondary's orbit.}
 \label{fig:plot3}
\end{figure*}

Comparing these four simulations, we can already distinguish a key difference in how the disc evolves depending on their EoS: if the internal energy of the disc cannot dissipate, the effective viscosity is enhanced, and therefore, the accretion rate too. This is because $\nu = \alpha_{\mathrm{ss}} c_s H = 0.1 \alpha^{\mathrm{av}}\langle h \rangle c_s$, and while in both cases the smoothing length is inversely proportional to the density, in the isothermal disc the sound speed is prescribed to be constant, while in the adiabatic scenario it increase over time since $c_s^2 \propto u$.

As we explained before, the thickness of the disc determines the efficiency of the tidal torque. Fixing this value in the simulation can define from the beginning how strong the coupling between the binary and the disc will be, while letting it vary over time will allow the system to go through different phases. For the pure adiabatic case, at $t = 2000$ the disc scale height has already increased an order of magnitude from the initial condition $H / R = 0.01$. As expected, this is independent of the presence of a secondary object in the inner region where the primary potential dominates, while at a larger radius, if the tidal effect exists, the scale height is enhanced until it saturates at a similar value through the whole disc.

\subsection{Simulations with added cooling}

As a next step, we include a cooling function in the adiabatic disc: we will let the gas dissipate its internal energy, which in turn will enable the disc to become thin or thick depending on the timescale of the process compared with the migration timescale of the secondary.

Figure \ref{fig:plot2} shows the accretion rate on to the primary black hole as a function of time for all the binary simulations in this work (3-11).
The top left panel is the general behaviour through all the simulation, while the bottom and right panels are zoom-in of relevant periods.

We can divide the overall evolution of the accretion in the isothermal case and the adiabatic with cooling ones into three main stages. In the beginning, the sharp initial condition produces an artificial super-Eddington accretion rate for $t \leq 100$, which affects the disc evolution after around $t \sim 1000$. After this transient behaviour, the accretion slows down reaching a quasi-steady state. Depending on how fast the disc can cool down, the constant value at which it converges and the time it takes to reach it varies: the longer the cooling timescale is, the higher is the accretion on to the primary black hole and it stabilizes faster. The scenario with $\beta = 40$ becomes constant before $t = 2000$ and maintains an Eddington accretion rate for most of the simulation until the majority of the disc is accreted and/or left behind the secondary just as the final stage begins. In this case, the final merger occurs in a gas-free environment, completely suppressing the precursor.

As we increase the efficiency of the cooling process, the accretion rate becomes lower and begins to present some transient features (see below), shown in the bottom-left panel of the same figure.
The decrease in the accretion can be seen from Equation \eqref{eq:alphass}, since $\dot{M} \propto \nu \propto \langle h \rangle (H / R)$: for a shorter cooling timescale, the disc becomes thinner and both the smoothing length $\langle h \rangle$ and the scale height $H / R$ are reduced.

\begin{figure}
 \includegraphics[width=.9\columnwidth]{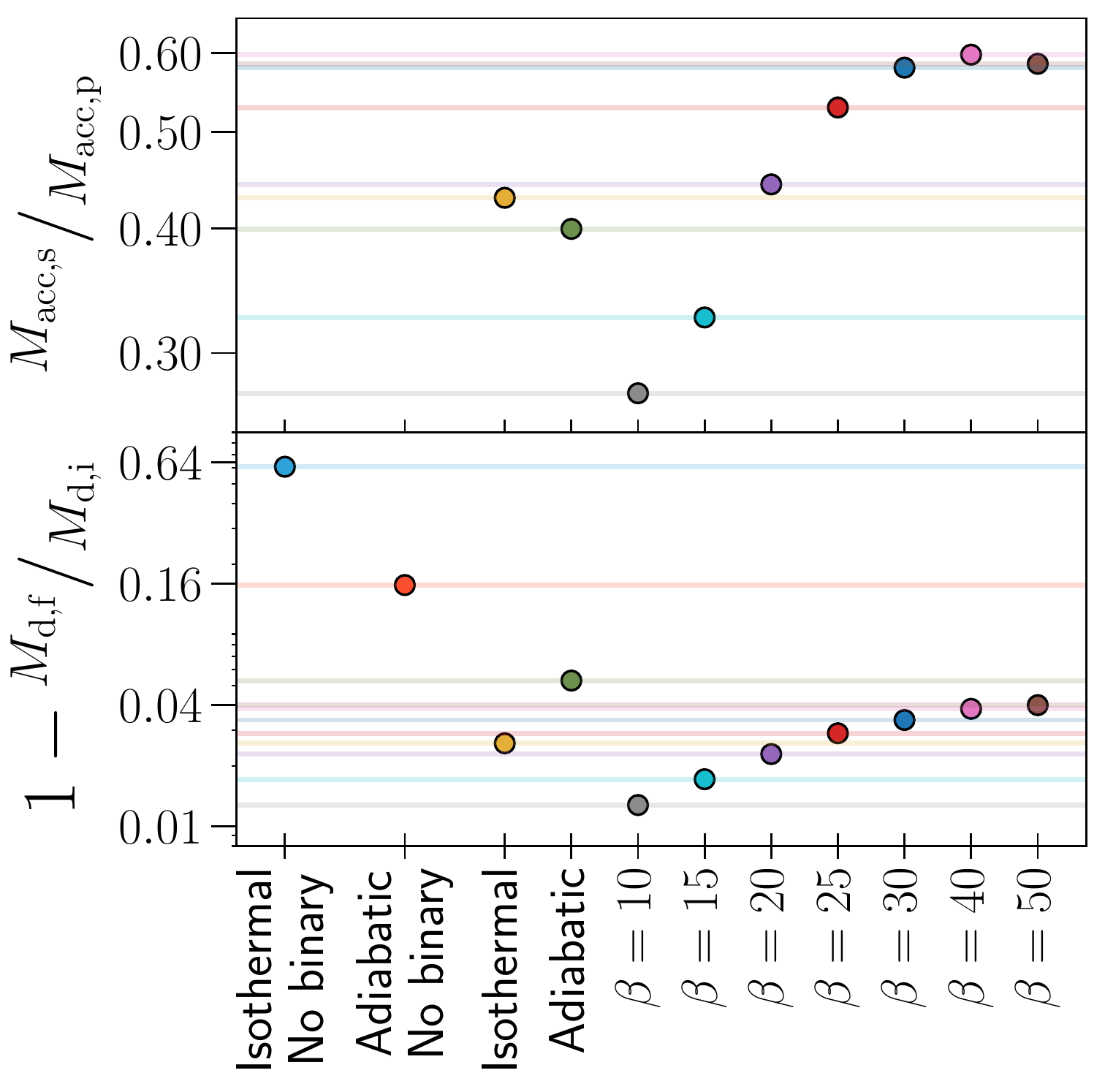}
 \caption{Top panel: Ratio between the total mass accreted by the secondary over the primary at the end of the simulation. The ratio increases as the cooling becomes less efficient, but at some point, the trend reverts and converges to the pure adiabatic case. 
 Bottom panel: Residual mass in the vicinity of the newly formed black hole. The cases without secondary retain more material than the rest since no tidal effect push it inwards. The pure adiabatic case with a secondary black hole is an upper limit of the final amount of gas left.}
 \label{fig:plot4}
\end{figure}

The final stage of the simulations is shown in the right panel of Figure \ref{fig:plot2}. We see that increasing the efficiency of the cooling process makes a stronger precursor. We expect this behaviour since a cooler disc is a thinner one, which increases the tidal influence from the secondary and decreases the viscous dissipation. For $\beta \geq 30$ the enhancement in the accretion rate is small, and we can consider this as the maximum value of the cooling parameter for which an electromagnetic counterpart could be produced. The simulations with $\beta = 25$ and $\beta = 20$ present a single peak around $t = 8500$, while the cases with a more efficient cooling (plus the isothermal case) show a second peak around $t = 9000$. As we decrease $\beta$, the dominance of the first peak shifts to the second one. 
The existence of a double peak and an earlier transient feature in models with low $\beta$ is due to Lindblad resonances reaching the accretion radius.  This is shown in detail in Appendix \ref{sec.app}, but we notice that general relativistic (GR) effects not included in the simulations would likely affect this result.

Figure \ref{fig:plot3} shows the surface density of the disc for five simulations, at four different times, including the initial condition.
A clear difference arises when we compare the first two columns. 
In the adiabatic model the gas retains its internal energy: as the temperature rises, the disc becomes thicker. 
The higher sound-speed and the decrease in the tidal influence of the smaller black hole allow the gas to move through the secondary's orbit and disperse in the medium. Already at $t = 6000$, there is almost no material inside the binary orbit, and as the rest goes away, it cools down and is pushed outward by the secondary.

The last three columns, where the discs are adiabatic but can gradually release their internal energy, behave similarly to the isothermal scenario. As we increase $\beta$, the inner disc looks less defined, and more material can leak away from the secondary's orbit. In all three cases, the outer shape of the disc is not circular, but rather slightly triangular due to the dominance of the $m = 3$ component of the binary potential since the timescale of the binary migration is much shorter than the viscous timescale. At $t = 9000$ $\beta = 10$ is the only case that presents a well distinct disc, even more, its shape is strongly distorted, again due to the $m = 3$ component of the gravitational potential (see Appendix \ref{sec.app}).

Finally, Figure \ref{fig:plot4} shows the fate of the initial gas in the disc at the end of each simulation. The top panel displays the ratio between the accreted mass on to the secondary over the primary, while the bottom panel shows the fraction of the material that survives the merger compared with the initial condition. 

In the adiabatic cases with cooling, the accretion on to the secondary is enhanced as the cooling timescale becomes longer since the disc thickens and the gas can approach the secondary black hole. Nevertheless, the accretion on to the secondary saturates at some $30 < \beta < 50$: after the material is hot enough to move on to the secondary's orbit, it can be accreted only if it encounters the black hole. While the dynamical timescale of the secondary is much shorter than the time the material needs to cross the orbit, so it will be easily accreted \citep[but see][]{Baruteau2012}, material could also move thought the orbit at a vertical distance larger than the size of the accretion radius of the secondary, allowing the gas to flow outwards without being accreted. 
Given the conditions of our simulation, the accretion rate on to the secondary should stall when $h / r \sim R_{\mathrm{acc,2}} / a$. Even more, since the available material should be similar in each case (once it can move to the secondary's orbit), and for longer cooling timescale the gas will cover a larger region, the final accretion on to the secondary should decrease. Regardless, as the binary shrinks, this scale height condition becomes bigger, so at the end of the simulation, any material between the black holes will be accreted by the primary or secondary.

The accretion on to the primary (not explicitly shown here) also has a changing trend: for smaller cooling timescale, the accretion is enhanced by the tidal effect, until $\beta \leq 20$, where is similar to the pure adiabatic scenario, then, for longer $\beta$ it becomes smaller. The reason for this change is the competition between the viscosity (enhanced for thicker discs) and the tidal effect 
(which in turn becomes inefficient), while the GW timescale constrains everything, as it determines how much material can be accreted on to the primary before the final stage of the simulation.

The bottom panel of the same figure shows the residual mass in the system after the merger. The first two points are the simulation without binary that we stop at the same time as when the merger should occur. 
The enhanced accretion of the pure adiabatic case without secondary makes the residual disc mass four times smaller than the single black hole isothermal case.
For the simulations with a secondary black hole, the pure adiabatic EoS defines an upper limit to the amount of material that survives the merger: as we reduce the cooling timescale, this residual mass becomes smaller. While for $\beta \leq 20$ the final amount of material in the black hole vicinity is mostly due to tidal effects and accretion onto the primary, for $20 \leq \beta \leq 40$ the secondary's accretion dominates. Finally, for $\beta \geq 40$,  the enhanced viscosity and accretion on to the primary lead again.

\section{Discussion}
\label{sec:caf}

There are several simplifications needed for the simulations presented here to be feasible: following a SMBH binary with $q \ll 1 $ decaying by GWs with a full 3D GR model of its disc system is currently impossible. 
\footnote{For state of the art on GR simulations of circumbinary discs see the reviews by \citet{Duez2019, Gold2019}.} 
For instance, \citet{Bowen2017} managed to follow an equal mass binary system from $a_0 = 20$ down to $a_f = 16$ in 2D. Since the migration time depends inversely on the mass ratio, our small secondary would prevent us from achieving even a fraction of that.

The main shortcoming in our work is the absence of relativistic effects, since they can alter the morphology and thermodynamics of the disc. 
For example, the ISCO should prevent material from orbiting at distances smaller than $6 R_{\rm{g}}$ from a non-rotating black hole. In our work, instead, the gas is allowed down to the accretion radius of the primary $R_{\rm{acc,p}} = 2$. In any case, \citet{Bowen2017} find in their $a_0 = 20$ simulation that the discs around each black hole exist inside the ISCO at least for the duration of their simulation, suggesting that our approach is not unrealistic. 

On the binary migration, as it shrinks, the velocity of the secondary will deviate from Peters' formula due to higher-order post-Newtonian effects. In our system of interest, since we are considering circular binaries at already close separations, this deviation is at most a factor of two \citep{Zwick2020}.

Another effect that we neglect is apsidal precession: this consequence of GR makes eccentric gas streams self-cross, which tends to circularize the orbits inner region \citep{Bonnerot2016}, smoothing out transient features for cooler discs, heating the gas and enhancing accretion. 
Still, more detailed work is needed to explore if the tidal distortion can become dominant, given the timescales of the different processes and their influence in the effective viscosity.

For the thermodynamics, the innermost region around the black hole is expected to be radiation pressure dominated, which would result in an effective $\gamma = 4/3$, making the fluid more compressible, while other effects (such as shocks due to apsidal precession, discussed above) could act as heating terms, enhancing accretion and thickness. Here, we use $\gamma = 5/3$ and consider that $\beta$ contains the deviations from the ideal gas scenario.

Finally, in order to characterize how our choice of parameters determines the behaviour of the system, we ran three tests taking simulation 7 ($\beta = 20$) as a reference. First, we duplicated the resolution and found no significant changes through all the simulation. Then we reduced the accretion radius of the secondary by a factor ten, allowing us to see spirals coming from the secondary black hole reported by \citet{Baruteau2012}. While this reduces the accretion onto the secondary and transfers the material outside the binary orbit, the primary accretion is unaffected. Figure \ref{fig:plot4} should change accordingly, varying the scale, but no the resulting trend as we change the value of $\beta$. In the last simulation, we used $\gamma = 4/3$, which is more appropriate for a radiation pressure dominated flow. This simulation looks almost identical to the case with $\gamma = 5/3$ and $\beta = 10$, showing that a smaller adiabatic index makes the disc thinner. Therefore, for our purposes, a change in the fluid adiabatic index merely results in a different effective cooling timescale, so that our results (that use a range of $\beta$) can be re-mapped easily for $\gamma=4/3$ by simply reducing $\beta$ accordingly.

Overall, the work presented here is one step further in the path to reach a realistic model of a decaying binary of SMBH. Instead of implementing currently-unaffordable GR effects, we have chosen to explore the effects of the thermodynamics, encoded in a range of effective cooling times. Future steps in this context could be the implementation of some of the missing GR effects in the simulation as recipes or using modified gravitational potentials.

\section{Conclusions}

In this work we studied the behaviour of an adiabatic inner disc in a binary of supermassive black holes at the end of the gravitational wave phase, complementing the work already done by \citet{Cerioli2016} and \citet{Pereira2019}. Our main result is the suppression of the precursor prior to the merger for discs with cooling were $\beta \geq 30$. For $\beta = 10$ the disc is thinner than the isothermal case for our choice of $H/R$, and the enhancement of the accretion at the final stage is almost two orders of magnitude compared with the rest of the simulations. The thickness of the discs obtained here are a lower limit since radiation pressure, which should dominate in this case, is not modelled.
While the general relativity effects should be relevant at this binary separation, we think they do not alter the main conclusion of our work, and the computational power needed to evolve a simulation with those characteristics become prohibitive.

The fact that our initial estimation of the cooling timescale yields $\beta \geq 100$ implies that the overall scenario, where material survives all the binary's evolution until this point, is challenging.  The presence of streams from the circumbinary disc can help feed up the inner disc and provide the seed of a precursor \citep{Bowen2019} if the disc can cool down efficiently enough. A new set of simulations, starting with a binary farther away, could help us understand if some material can survive.

The secondary black hole accretion disc, not modelled in this work, can be a promising alternative source of a periodic electromagnetic precursor. As pointed out by \citet{Cerioli2016}, the secondary also manifests a weak precursor for $\beta \leq 20$. For binaries with larger mass ratios, the smaller companion will receive more gas from the streams coming through the cavity \citep{Dunhill2015}. This new gas could help sustain a mini disc, which in the GW stage will also be squeezed.

\section*{Acknowledgment}

 This project has received funding from the European Union's Horizon 2020 research and innovation programme under the Marie Sk\l{}odowska-Curie grant agreement No 823823 (DUSTBUSTERS).
 CF acknowledges financial support from ANID through PFCHA/Doctorado Nacional (2017-21171063) and Basal (AFB-170002).
 CF thanks Universit\`a degli studi di Milano for the warm hospitality during his visit, where this work started.
 CF and JC acknowledge support from Iniciativa Cient\'ifica Milenio via the N\'ucleo Milenio de Formaci\'on Planetaria.  
 The Geryon2 cluster housed at the Centro de Astro-Ingenier\'ia UC was used for the calculations performed in this paper. The BASAL PFB-06 CATA, Anillo ACT-86, FONDEQUIP AIC-57, and QUIMAL 130008 provided funding for several improvements to the Geryon/Geryon2 cluster.

\bibliographystyle{mnras}

\begin{thebibliography}{}
\makeatletter
\relax
\def\mn@urlcharsother{\let\do\@makeother \do\$\do\&\do\#\do\^\do\_\do\%\do\~}
\def\mn@doi{\begingroup\mn@urlcharsother \@ifnextchar [ {\mn@doi@}
 {\mn@doi@[]}}
\def\mn@doi@[#1]#2{\def\@tempa{#1}\ifx\@tempa\@empty \href
 {http://dx.doi.org/#2} {doi:#2}\else \href {http://dx.doi.org/#2} {#1}\fi
 \endgroup}
\def\mn@eprint#1#2{\mn@eprint@#1:#2::\@nil}
\def\mn@eprint@arXiv#1{\href {http://arxiv.org/abs/#1} {{\tt arXiv:#1}}}
\def\mn@eprint@dblp#1{\href {http://dblp.uni-trier.de/rec/bibtex/#1.xml}
 {dblp:#1}}
\def\mn@eprint@#1:#2:#3:#4\@nil{\def\@tempa {#1}\def\@tempb {#2}\def\@tempc
 {#3}\ifx \@tempc \@empty \let \@tempc \@tempb \let \@tempb \@tempa \fi \ifx
 \@tempb \@empty \def\@tempb {arXiv}\fi \@ifundefined
 {mn@eprint@\@tempb}{\@tempb:\@tempc}{\expandafter \expandafter \csname
 mn@eprint@\@tempb\endcsname \expandafter{\@tempc}}}
 
\bibitem[\protect\citeauthoryear
    {Abbott et al.}
    {Abbott et al.}{2016}]
    {Abbott2016}
    Abbott B.~P., Abbott R., Abbott T.~D., et al., 2016, \mn@doi [PhRvL]
    {10.1103/PhysRevLett.116.061102 }
    \href {https://ui.adsabs.harvard.edu/abs/2016PhRvL.116f1102A},
{116, 061102}

\bibitem[\protect\citeauthoryear
    {Armitage \& Natarajan}
    {Armitage \& Natarajan}{2002}]
    {Armitage2002}
    Armitage P.~J., Natarajan P., 2002, \mn@doi [ApJL]
    {10.1086/339770 }
    \href {https://ui.adsabs.harvard.edu/abs/2002ApJ...567L...9A},
{567, L9}

\bibitem[\protect\citeauthoryear
    {Artymowicz \& Lubow}
    {Artymowicz \& Lubow}{1994}]
    {Artymowicz1994}
    Artymowicz P., Lubow S.~H., 1994, \mn@doi [ApJ]
    {10.1086/173679 }
    \href {https://ui.adsabs.harvard.edu/abs/1994ApJ...421..651A},
{421, 651}

\bibitem[\protect\citeauthoryear
    {Baruteau, Ramirez-Ruiz, \& Masset}
    {Baruteau et al.}{2012}]
    {Baruteau2012}
    Baruteau C., Ramirez-Ruiz E., Masset F., 2012, \mn@doi [MNRAS]
    {10.1111/j.1745-3933.2012.01258.x }
    \href {https://ui.adsabs.harvard.edu/abs/2012MNRAS.423L..65B},
{423, L65}

\bibitem[\protect\citeauthoryear
    {Begelman, Blandford, \& Rees}
    {Begelman et al.}{1980}]
    {Begelman1980}
    Begelman M.~C., Blandford R.~D., Rees M.~J., 1980, \mn@doi [Natur]
    {10.1038/287307a0 }
    \href {https://ui.adsabs.harvard.edu/abs/1980Natur.287..307B},
{287, 307}

\bibitem[\protect\citeauthoryear
    {Bonnerot et al.}
    {Bonnerot et al.}{2016}]
    {Bonnerot2016}
    Bonnerot C., Rossi E., Lodato G., Price D., 2016, \mn@doi [MNRAS]
    {10.1093/mnras/stv2411}
    \href {https://ui.adsabs.harvard.edu/abs/2016MNRAS.455.2253B},
{455, 2253}

\bibitem[\protect\citeauthoryear
	{Bowen et al.}
	{Bowen et al.}{2017}]
	{Bowen2017}
	Bowen D.~B., Campanelli M., Krolik J.~H., Mewes V., Noble S.~C., 2017, \mn@doi [ApJ]
	{10.3847/1538-4357/aa63f3 }
	\href {https://ui.adsabs.harvard.edu/abs/2017ApJ...838...42B},
{838, 42}


\bibitem[\protect\citeauthoryear
    {Bowen et al.}
    {Bowen et al.}{2019}]
    {Bowen2019}
    Bowen D.~B., Mewes V., Noble S.~C., Avara M., Campanelli M., Krolik J.~H., 2019, \mn@doi [ApJ]
    {10.3847/1538-4357/ab2453 }
    \href {https://ui.adsabs.harvard.edu/abs/2019ApJ...879...76B},
{879, 76}

\bibitem[\protect\citeauthoryear
    {Centrella et al.}
    {Centrella et al.}{2010}]
    {Centrella2010}
    Centrella J., Baker J.~G., Kelly B.~J., van Meter J.~R., 2010, \mn@doi [RvMP]
    {10.1103/RevModPhys.82.3069 }
    \href {https://ui.adsabs.harvard.edu/abs/2010RvMP...82.3069C},
{82, 3069}

\bibitem[\protect\citeauthoryear
    {Cerioli, Lodato, \& Price}
    {Cerioli et al.}{2016}]
    {Cerioli2016}
    Cerioli A., Lodato G., Price D.~J., 2016, \mn@doi [MNRAS]
    {10.1093/mnras/stw034 }
    \href {https://ui.adsabs.harvard.edu/abs/2016MNRAS.457..939C},
{457, 939}

\bibitem[\protect\citeauthoryear
    {Chang et al.}
    {Chang et al.}{2010}]
    {Chang2010}
    Chang P., Strubbe L.~E., Menou K., Quataert E., 2010, \mn@doi [MNRAS]
    {10.1111/j.1365-2966.2010.17056.x }
    \href {https://ui.adsabs.harvard.edu/abs/2010MNRAS.407.2007C},
{407, 2007}


\bibitem[\protect\citeauthoryear
    {Colpi \& Dotti}
    {Colpi \& Dotti}{2011}]
    {Colpi2011}
    Colpi M., Dotti M., 2011, \mn@doi [ASL]
    {10.1166/asl.2011.1205 }
    \href {https://ui.adsabs.harvard.edu/abs/2011ASL.....4..181C},
{4, 181}

\bibitem[\protect\citeauthoryear
    {Cuadra et al.}
    {Cuadra et al.}{2009}]
    {Cuadra2009}
    Cuadra J., Armitage P.~J., Alexander R.~D., Begelman M.~C., 2009, \mn@doi [MNRAS]
    {10.1111/j.1365-2966.2008.14147.x }
    \href {https://ui.adsabs.harvard.edu/abs/2009MNRAS.393.1423C},
{393, 1423}

\bibitem[\protect\citeauthoryear
    {D'Orazio et al.}
    {D'Orazio et al.}{2016}]
    {DOrazio2016}
    D'Orazio D.~J., Haiman Z., Duffell P., MacFadyen A., Farris B., 2016, \mn@doi [MNRAS]
    {10.1093/mnras/stw792 }
    \href {https://ui.adsabs.harvard.edu/abs/2016MNRAS.459.2379D},
{459, 2379}

\bibitem[\protect\citeauthoryear{Duez \& Zlochower}{2019}]{Duez2019} Duez M.~D., Zlochower Y., 2019, RPPh, 82, 016902


\bibitem[\protect\citeauthoryear
    {Dunhill, Cuadra, \& Dougados}
    {Dunhill et al.}{2015}]
    {Dunhill2015}
    Dunhill A.~C., Cuadra J., Dougados C., 2015, \mn@doi [MNRAS]
    {10.1093/mnras/stv284 }
    \href {https://ui.adsabs.harvard.edu/abs/2015MNRAS.448.3545D},
{448, 3545}

\bibitem[\protect\citeauthoryear
    {eLISA Consortium et al.}
    {eLISA Consortium et al.}{2013}]
    {eLISA2013}
    eLISA Consortium, Amaro Seoane P., Aoudia S., et al., 2013, \mn@doi [arXiv]
    { }
    \href {https://ui.adsabs.harvard.edu/abs/2013arXiv1305.5720E},
{ arXiv:1305.5720}

\bibitem[\protect\citeauthoryear
    {Escala et al.}
    {Escala et al.}{2005}]
    {Escala2005}
    Escala A., Larson R.~B., Coppi P.~S., Mardones D., 2005, \mn@doi [ApJ]
    {10.1086/431747 }
    \href {https://ui.adsabs.harvard.edu/abs/2005ApJ...630..152E},
{630, 152}

\bibitem[\protect\citeauthoryear
    {Ferrarese \& Merritt}
    {Ferrarese \& Merritt}{2000}]
    {Ferrarese2000}
    Ferrarese L., Merritt D., 2000, \mn@doi [ApJL]
    {10.1086/312838 }
    \href {https://ui.adsabs.harvard.edu/abs/2000ApJ...539L...9F},
{539, L9}

\bibitem[\protect\citeauthoryear
    {Fontecilla, Chen, \& Cuadra}
    {Fontecilla et al.}{2017}]
    {Fontecilla2017}
    Fontecilla C., Chen X., Cuadra J., 2017, \mn@doi [MNRAS]
    {10.1093/mnrasl/slw258 }
    \href {https://ui.adsabs.harvard.edu/abs/2017MNRAS.468L..50F},
{468, L50}

\bibitem[\protect\citeauthoryear
    {Fontecilla, Haiman, \& Cuadra}
    {Fontecilla et al.}{2019}]
    {Fontecilla2019}
    Fontecilla C., Haiman Z., Cuadra J., 2019, \mn@doi [MNRAS]
    {10.1093/mnras/sty2972 }
    \href {https://ui.adsabs.harvard.edu/abs/2019MNRAS.482.4383F},
{482, 4383}

\bibitem[\protect\citeauthoryear
    {Frank, King, \& Raine}
    {Frank et al.}{2002}]
    {Frank2002}
    Frank J., King A., Raine D.~J., 2002, \mn@doi [apa..book]
    { }
    \href {https://ui.adsabs.harvard.edu/abs/2002apa..book.....F},
{ 398}

\bibitem[\protect\citeauthoryear
    {Gammie}
    {Gammie}{2001}]
    {Gammie2001}
    Gammie C.~F., 2001, \mn@doi [ApJ]
    {10.1086/320631 }
    \href {https://ui.adsabs.harvard.edu/abs/2001ApJ...553..174G},
{553, 174}


\bibitem[\protect\citeauthoryear{Goicovic et al.}{2017}]{Goicovic2017} Goicovic F.~G., Sesana A., Cuadra J., Stasyszyn F., 2017, MNRAS, 472, 514


\bibitem[\protect\citeauthoryear
	{Gold et al.}
	{Gold et al.}{2014}]
	{Gold2014}
	Gold R., Paschalidis V., Ruiz M., Shapiro S.~L., Etienne Z.~B., Pfeiffer H.~P., 2014, \mn@doi [PhRvD]
	{10.1103/PhysRevD.90.104030 }
	\href {https://ui.adsabs.harvard.edu/abs/2014PhRvD..90j4030G},
{90, 104030}

\bibitem[\protect\citeauthoryear{Gold}{2019}]{Gold2019} Gold, 2019, Galax, 7, 63


\bibitem[\protect\citeauthoryear
    {Governato, Colpi, \& Maraschi}
    {Governato et al.}{1994}]
    {Governato1994}
    Governato F., Colpi M., Maraschi L., 1994, \mn@doi [MNRAS]
    {10.1093/mnras/271.2.317 }
    \href {https://ui.adsabs.harvard.edu/abs/1994MNRAS.271..317G},
{271, 317}

\bibitem[\protect\citeauthoryear
    {Kocsis \& Loeb}
    {Kocsis \& Loeb}{2008}]
    {Kocsis2008}
    Kocsis B., Loeb A., 2008, \mn@doi [PhRvL]
    {10.1103/PhysRevLett.101.041101 }
    \href {https://ui.adsabs.harvard.edu/abs/2008PhRvL.101d1101K},
{101, 041101}

\bibitem[\protect\citeauthoryear
    {Kocsis, Haiman, \& Loeb}
    {Kocsis et al.}{2012}]
    {Kocsis2012}
    Kocsis B., Haiman Z., Loeb A., 2012, \mn@doi [MNRAS]
    {10.1111/j.1365-2966.2012.22118.x }
    \href {https://ui.adsabs.harvard.edu/abs/2012MNRAS.427.2680K},
{427, 2680}

\bibitem[\protect\citeauthoryear
    {Kormendy \& Ho}
    {Kormendy \& Ho}{2013}]
    {Kormendy2013}
    Kormendy J., Ho L.~C., 2013, \mn@doi [ARA\&A]
    {10.1146/annurev-astro-082708-101811 }
    \href {https://ui.adsabs.harvard.edu/abs/2013ARA\&A..51..511K},
{51, 511}

\bibitem[\protect\citeauthoryear
    {Lin \& Papaloizou}
    {Lin \& Papaloizou}{1979}]
    {Lin1979}
    Lin D.~N.~C., Papaloizou J., 1979, \mn@doi [MNRAS]
    {10.1093/mnras/186.4.799 }
    \href {https://ui.adsabs.harvard.edu/abs/1979MNRAS.186..799L},
{186, 799}

\bibitem[\protect\citeauthoryear
    {Lin \& Papaloizou}
    {Lin \& Papaloizou}{1986}]
    {Lin1986}
    Lin D.~N.~C., Papaloizou J., 1986, \mn@doi [ApJ]
    {10.1086/164653 }
    \href {https://ui.adsabs.harvard.edu/abs/1986ApJ...309..846L},
{309, 846}

\bibitem[\protect\citeauthoryear
    {Lodato et al.}
    {Lodato et al.}{2009}]
    {Lodato2009}
    Lodato G., Nayakshin S., King A.~R., Pringle J.~E., 2009, \mn@doi [MNRAS]
    {10.1111/j.1365-2966.2009.15179.x }
    \href {https://ui.adsabs.harvard.edu/abs/2009MNRAS.398.1392L},
{398, 1392}

\bibitem[\protect\citeauthoryear
    {Mayer \& Bonoli}
    {Mayer \& Bonoli}{2019}]
    {Mayer2019}
    Mayer L., Bonoli S., 2019, \mn@doi [RPPh]
    {10.1088/1361-6633/aad6a5 }
    \href {https://ui.adsabs.harvard.edu/abs/2019RPPh...82a6901M},
{82, 016901}

\bibitem[\protect\citeauthoryear
    {Merritt \& Milosavljevi{\'c}}
    {Merritt \& Milosavljevi{\'c}}{2005}]
    {Merritt2005}
    Merritt D., Milosavljevi{\'c} M., 2005, \mn@doi [LRR]
    {10.12942/lrr-2005-8 }
    \href {https://ui.adsabs.harvard.edu/abs/2005LRR.....8....8M},
{8, 8}

\bibitem[\protect\citeauthoryear
    {Milosavljevi{\'c} \& Merritt}
    {Milosavljevi{\'c} \& Merritt}{2001}]
    {Milosavljevic2001}
    Milosavljevi{\'c} M., Merritt D., 2001, \mn@doi [ApJ]
    {10.1086/323830 }
    \href {https://ui.adsabs.harvard.edu/abs/2001ApJ...563...34M},
{563, 34}

\bibitem[\protect\citeauthoryear
    {Milosavljevi{\'c} \& Merritt}
    {Milosavljevi{\'c} \& Merritt}{2003}]
    {Milosavljevic2003}
    Milosavljevi{\'c} M., Merritt D., 2003, \mn@doi [ApJ]
    {10.1086/378086 }
    \href {https://ui.adsabs.harvard.edu/abs/2003ApJ...596..860M},
{596, 860}

\bibitem[\protect\citeauthoryear
    {Pereira et al.}
    {Pereira et al.}{2019}]
    {Pereira2019}
    Pereira F.~A.~C., Lodato G., Rodrigues I., Alves M.~E.~S., Price D.~J., 2019, \mn@doi [MNRAS]
    {10.1093/mnras/sty3471 }
    \href {https://ui.adsabs.harvard.edu/abs/2019MNRAS.484...31P},
{484, 31}

\bibitem[\protect\citeauthoryear
    {Peters}
    {Peters}{1964}]
    {Peters1964}
    Peters P.~C., 1964, \mn@doi [PhRv]
    {10.1103/PhysRev.136.B1224 }
    \href {https://ui.adsabs.harvard.edu/abs/1964PhRv..136.1224P},
{136, 1224}

\bibitem[\protect\citeauthoryear
    {Price et al.}
    {Price et al.}{2018}]
    {Price2018}
    Price D.~J., Wurster J., Tricco T.~S., et al., 2018, \mn@doi [PASA]
    {10.1017/pasa.2018.25 }
    \href {https://ui.adsabs.harvard.edu/abs/2018PASA...35...31P},
{35, e031}

\bibitem[\protect\citeauthoryear
    {Ragusa, Lodato, \& Price}
    {Ragusa et al.}{2016}]
    {Ragusa2016}
    Ragusa E., Lodato G., Price D.~J., 2016, \mn@doi [MNRAS]
    {10.1093/mnras/stw1081 }
    \href {https://ui.adsabs.harvard.edu/abs/2016MNRAS.460.1243R},
{460, 1243}

\bibitem[\protect\citeauthoryear
    {Reines et al.}
    {Reines et al.}{2014}]
    {Reines2014}
    Reines A.~E., Plotkin R.~M., Russell T.~D., Mezcua M., Condon J.~J., Sivakoff G.~R., Johnson K.~E., 2014, \mn@doi [ApJL]
    {10.1088/2041-8205/787/2/L30 }
    \href {https://ui.adsabs.harvard.edu/abs/2014ApJ...787L..30R},
{787, L30}

\bibitem[\protect\citeauthoryear
    {Roedig et al.}
    {Roedig et al.}{2011}]
    {Roedig2011}
    Roedig C., Dotti M., Sesana A., Cuadra J., Colpi M., 2011, \mn@doi [MNRAS]
    {10.1111/j.1365-2966.2011.18927.x }
    \href {https://ui.adsabs.harvard.edu/abs/2011MNRAS.415.3033R},
{415, 3033}

\bibitem[\protect\citeauthoryear
    {Shakura \& Sunyaev}
    {Shakura \& Sunyaev}{1973}]
    {Shakura1973}
    Shakura N.~I., Sunyaev R.~A., 1973, \mn@doi [A\&A]
    { }
    \href {https://ui.adsabs.harvard.edu/abs/1973A\&A....24..337S},
{500, 33}

\bibitem[\protect\citeauthoryear
    {Shannon et al.}
    {Shannon et al.}{2015}]
    {Shannon2015}
    Shannon R.~M., Ravi V., Lentati L.~T., et al., 2015, \mn@doi [Sci]
    {10.1126/science.aab1910 }
    \href {https://ui.adsabs.harvard.edu/abs/2015Sci...349.1522S},
{349, 1522}


\bibitem[\protect\citeauthoryear
    {Tazzari \& Lodato}
    {Tazzari \& Lodato}{2015}]
    {Tazzari2015}
    Tazzari M., Lodato G., 2015, \mn@doi [MNRAS]
    {10.1093/mnras/stv352 }
    \href {https://ui.adsabs.harvard.edu/abs/2015MNRAS.449.1118T},
{449, 1118}

\bibitem[\protect\citeauthoryear
    {Volonteri, Haardt, \& Madau}
    {Volonteri et al.}{2003}]
    {Volonteri2003}
    Volonteri M., Haardt F., Madau P., 2003, \mn@doi [ApJ]
    {10.1086/344675 }
    \href {https://ui.adsabs.harvard.edu/abs/2003ApJ...582..559V},
{582, 559}

\bibitem[\protect\citeauthoryear
    {Yu}
    {Yu}{2002}]
    {Yu2002}
    Yu Q., 2002, \mn@doi [MNRAS]
    {10.1046/j.1365-8711.2002.05242.x }
    \href {https://ui.adsabs.harvard.edu/abs/2002MNRAS.331..935Y},
{331, 935}

\bibitem[\protect\citeauthoryear
	{Zwick et al.}
	{Zwick et al.}{2020}]
	{Zwick2020}
	Zwick L., Capelo P.~R., Bortolas E., Mayer L., Amaro-Seoane P., 2020, \mn@doi [MNRAS]
	{10.1093/mnras/staa1314 }
	\href {https://ui.adsabs.harvard.edu/abs/2020MNRAS.495.2321Z},
{495, 2321}

\end{thebibliography}

\newpage

\appendix
\section{Accretion due to Lindblad resonance migration}
\label{sec.app}
The position of the inner Lindblad resonances, where the transfer of angular momentum from the binary on to the disc is enhanced and spiral arms generate, depends on the separation between primary and secondary and, as this distance shrinks due to the gravitational wave emission, the position of the resonances shift inwards.
The location of the $m$ mode inner Lindblad resonance is 
\begin{equation}
  r_{\mathrm{l,in}} = \left(1 - \frac{1}{m}\right)^{2/3} a,
\end{equation}
so, for $m = 2, r_{\mathrm{l,in}} = 0.63a$ and $m = 3, r_{\mathrm{l,in}} = 0.76a$. These two resonances are the ones that dominate in our work.

\begin{figure}
 \includegraphics[width=.9\columnwidth]{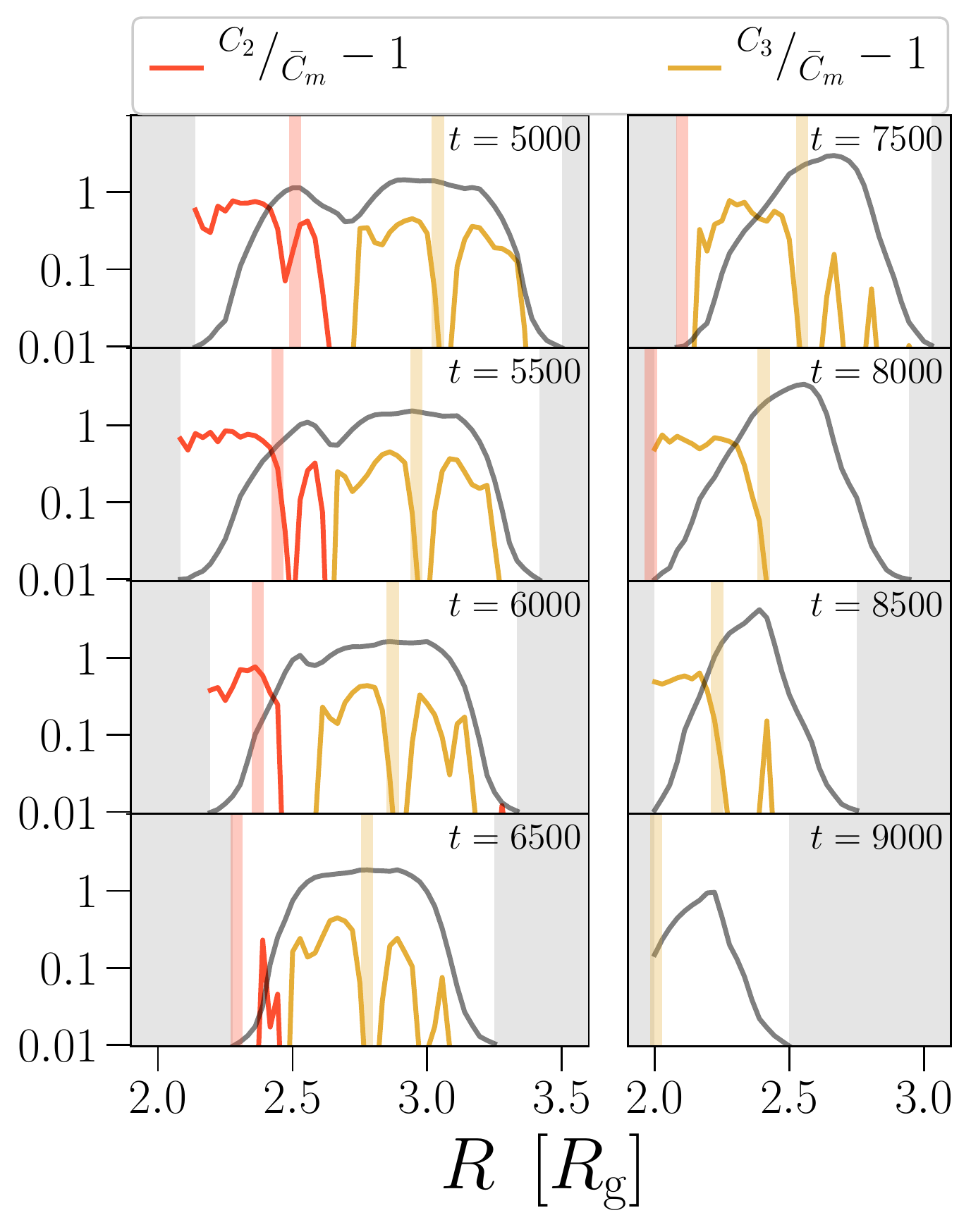}
 \caption{Relative strength of the $m = 2$ (red) and $m = 3$ (yellow) modes of the Fourier transform of the disc angular distribution as a function of  radius, for different snapshots of the simulation with $\beta = 10$. The vertical lines of the same colour show the position of the corresponding Lindblad resonance, which moves inwards due to the secondary's migration. The light gray region is outside the inner disc, while the dark gray line is the normalized surface density shown for reference.}
 \label{fig:plot5}
\end{figure}

When the binary goes from $a \sim 4.17$ to $ a \sim 3.6$, the location of the resonance that produce the $m=2$ spiral arm moves from $r \sim 2.45$ to $r \sim 2.3$; this produces a transient feature around $ 4000 \lesssim t \lesssim 6500$, when a overdensity tied to this resonance leaves the inner region of the disc and finally enters the accretion radius of the primary. At this final position, the local surface density is less than one per cent of the average density inside the disc. We confirm this hypothesis by calculating the Fourier transform of the SPH gas particles angular distribution at each snapshot and compare the strength of each mode,
\begin{equation}
  C_m (R)= \frac{1}{(2 \pi)^2} \frac{M_\mathrm{ring}(R)}{M_d} \left|\sum_{k=1}^j e^{-i m\theta_k}\right|
\end{equation}
where the sum is over all the $j$ particles in a given ring and $M_\mathrm{ring}(R) = j \times m_\mathrm{SPH}$, with $m_\mathrm{SPH}$ the mass of each particle, which is fixed in our simulations.
The right panels in Figure \ref{fig:plot5} show how the relative strength of the $m=2$ and $m=3$ modes evolve through this period for the case with $\beta = 10$.

The second peak for cooler discs, seen at the right of Figure \ref{fig:plot3} coincides with the binary separation where the $m=3$ Lindblad resonance leaves the inner region of the disc, shown in the right panels of Figure \ref{fig:plot5}. This is because a more efficient cooling depletes the region of the $m=2$ resonance, making the $m=3$ dominate. When the radius of this resonance leaves the disc, material in the spiral arm is dragged into the SMBH, enhancing the accretion.


\bsp    
\label{lastpage}
\end{document}